# A Dynamic Coupling Model of Optical Conductivity in Mixed-Valence Systems

Kin Yip Wong

## ABSTRACT

Using Linear Response Theory [17], with appropriate wave functions and energies from perturbation method, the absorption profiles can be calculated for all three classes of mixed-valence systems as defined by Robin and Day [37]: Class III (delocalized), Class I (localized) and Class II (intermediate between III and I). Based on these absorption profiles, one can calculate the corresponding frequency-dependent optical conductivity $\sigma(\omega)$ profiles with the following results:

(1) For all three classes, their $\sigma(\omega)$s' profiles are similar to their corresponding absorption profiles in regard to band shape and polarization, except peaks of $\sigma(\omega)$ profiles' tend to shift toward higher frequency with respect to the absorption profiles.

(2) The charge transfer absorption (CT band) is the major contributors of $\sigma(\omega)$. Moreover, the CT-induced IR band, also contributes to $\sigma(\omega)$, as it borrows its intensity from the CT band and the amount of borrowing depends on its proximity to the latter, as in Class II.

## INTRODUCTION

It is well known that mixed valence compounds exhibit various degree of electrical conductivities and have been studied both experimentally [1-8] and theoretically [9-15]. However, the above studies mostly focus on the DC conductivities and ignore the AC or optical conductivities $\sigma(\omega)$, which is frequency dependent. In the following sections, the CT and IR absorption bands are first calculated for each of the mixed-valence Classes using Linear Response Theory. Their optical conductivities $\sigma(\omega)$ are then calculated.

### PKS Vibronic Model in Operator Form

In a previous paper [16], it is shown that the Hamiltonian in the PKS model describing a dimer system in which an electron can transfer from one subunit to another (for a symmetric system):

$$H = \sum_i W_i^0 + \sum_{\alpha,i} P_i^{\alpha 2}/2M_i + \sum_{\alpha,i} \tfrac{1}{2} k_\alpha Q_i^{\alpha 2} + \sum_{\alpha,i} l_\alpha Q_i^\alpha + V_{12} \qquad (1)$$

(1) can be re-written in an operator form as follows:

$$H_{vib} = \sum_\alpha \hbar w_\alpha [\tfrac{1}{4}(\dot{q}_-^{\alpha 2} + q_-^{\alpha 2}) + 1/\sqrt{2}\,\lambda_\alpha q_-^\alpha \delta_n] + \varepsilon(a_1^+ a_2^- + a_1^- a_2^+) \qquad (2)$$

Where

$q_{\pm}^{\alpha} = 1/\sqrt{2}\,(q_1^{\alpha} \pm q_2^{\alpha})$ are the dimeric normal coordinates and $\delta_n = (n_1 - n_2)$.

It is seen that only the anti-symmetry mode $\boldsymbol{q_{-}^{\alpha}}$ interacts with the $\boldsymbol{\delta_n}$ term, which is related to the charge-transfer interaction.

In the presence of an electromagnetic field **F**, the total Hamiltonian can be written as

$$H_{total} = \sum_{\alpha} \hbar w_{\alpha} \left[ 1/4\,(\dot{q}_{-}^{\alpha 2}) + \tfrac{1}{\sqrt{2}} \lambda_{\alpha} q_{-}^{\alpha} \delta_n \right] +$$
$$\varepsilon(a_1^+ a_2^- + a_1^- a_2^+) - 1/2\,(e{\cdot}a){\cdot}(n_1 - n_2){\cdot}F \tag{3}$$

where **F** denotes an externally applied frequency-dependent electric field. The term $(e{\cdot}a)(n_1 - n_2)$ represents the electric dipole moment of the dimer, and the vector length **a** is of the order of the separation between the subunits. Equation (3) is identical to that for a two-site, single-electron organic conductor system described by Rice et al. earlier. The frequency-dependent conductivity (called scalar conductivity by Rice) is thus given by [17].

$$\sigma(w) = -iw\left(\tfrac{1}{4}e^2 \boldsymbol{a} \cdot \boldsymbol{a}\right)\left[ X(w) + \frac{X(w)\overline{X}(w)D(w)}{1 - \overline{X}(w)D(w)} \right] \tag{4}$$

where $\overline{X}(w) = X(w)/X(0)$ is defined as

$$X(w) = \sum_{\beta} |\langle \beta | \delta_n | 0 \rangle|^2\, 2w_{\beta 0}/\left[w_{\beta 0}^2 - (w + i\delta)^2\right] \tag{5}$$

where $|\beta\rangle$ denotes the eigenstate. $\beta = 0$ denotes the ground state. $D(w)$ is defined as

$$D(w) = \sum_{\alpha} X(0) g_{\alpha}^2\, w_{\alpha}/(w_{\alpha}^2 - w^2) \tag{6}$$

Note that in expression (4) the real part of the conductivity $\sigma(\omega)$ gives the charge-transfer absorption (i.e., CT band). From (4) it can be seen that there are two types of transitions. One is given by (5) and is centered at $w_{\beta 0}$ and is the charge-transfer or inter-valence transition due to electron transfer from one subunit to the other. Notice that $w \approx w_{\beta 0} \gg w_{\alpha}$. The second transition given by the second term in the square brackets in (4) is in the vibrational energy region (mid- to far-IR or the IR band). It is seen to be proportional to $D(w)$ and $X(w)$. Hence its intensities are derived from the charge-transfer transition and should therefore have the same polarization properties as the charge-transfer transition. From (6) it is readily seen that the mid-IR transition intensity is proportional to $g_{\alpha}^2$ (where $g_{\alpha} = \lambda_{\alpha} \hbar w_{\alpha}$) which is a measure of the electron-phonon coupling.

Note also that $\sigma(w)$ is proportional to $w$, and this implies that $\sigma(w)$ tends to zero as $w$ tends to 0.

**DELOCALIZED – CLASS III ($|\varepsilon_0| \gg \lambda_0^2$)**

The Hamiltonian in (2) can be solved using perturbation theory by defining a new basis set as follows:

$$|+\rangle = \tfrac{1}{\sqrt{2}}(|1\rangle + |2\rangle) \qquad |-\rangle = \tfrac{1}{\sqrt{2}}(|1\rangle - |2\rangle) \tag{7}$$

We get

$$H_{vib}/\hbar w_\alpha = \begin{bmatrix} \tfrac{1}{2}(\dot{q}_0^2 + q_0^2) + \varepsilon_0 & \lambda_0 q_0 \\ \lambda_0 q_0 & \tfrac{1}{2}(\dot{q}_0^2 + q_0^2) - \varepsilon_0 \end{bmatrix} \tag{8}$$

where $|1\rangle = a_1^+|0\rangle$, $|2\rangle = a_2^+|0\rangle$, and $|0\rangle$ is some zero state.

Eq. (8) is identical with that in the PKS model [18]. Note that $\lambda_0$ *and* $\varepsilon_0$ *are respectively related to the reorganisation energy and the resonance energy* [34,35] as follows:

Resonance Energy $\boldsymbol{H_{ab}} = \varepsilon_0/\hbar w_\alpha$
Reorganization Energy $\boldsymbol{\lambda} = 2\lambda_0^2/\hbar w_\alpha$

Note that $|\varepsilon_0| \gg \lambda_0^2$ is the condition: 2* Resonance Energy >> Reorganization Energy for a Delocalized Class III system as derived by Robin & Day [37].

The off-diagonal term $\lambda_0 q_0$ acts as a perturbation to the vibronic Hamiltonian $H_{vib}$. The zeroth-order wave function is [20, 22]:

$$\left|\mu_\pm^{(0)}\right\rangle = \tfrac{1}{\sqrt{2}}[|1\rangle \pm (-1)^\mu |2\rangle] * \chi_\mu(q_0) \tag{9}$$

with the corresponding zeroth-order energy [20, 22]:

$$E_\pm^{(0)}(\mu) = \mu + \tfrac{1}{2} \pm (-1)^\mu |\varepsilon_0| \tag{10}$$

The perturbation $\lambda_0 q_0$ mixes the vibrational states according to:

$$\lambda_0 q_0 \chi_\mu(q_0) = \tfrac{1}{\sqrt{2}} \lambda_0 [\mu^{1/2} \chi_{\mu-1}(q_0) + (\mu+1)^{1/2} \chi_{\mu+1}(q_0)] \tag{11}$$

Thus the first-order wave function is:

$$\left|\mu_\pm^{(1)}\right\rangle = \left|\mu_\pm^{(0)}\right\rangle + \tfrac{1}{\sqrt{2}} \lambda_0 \begin{bmatrix} \mu^{\tfrac{1}{2}} \left|(\mu-1)_\pm^{(0)}\right\rangle / (1 \mp 2(-1)^\mu |\varepsilon_0|) + \\ (\mu+1)^{\tfrac{1}{2}} \left|(\mu+1)_\pm^{(0)}\right\rangle / (-1 \mp 2(-1)^\mu |\varepsilon_0|) \end{bmatrix} \tag{12}$$

With use of (12) the transition matrix element or dipole moment can be evaluated [32]:

$$\left\langle v_+^{(1)}\middle|\delta n\middle|\mu_-^{(1)}\right\rangle = \left[1 + \tfrac{1}{2}\lambda_0^2 \left(\frac{(\mu v)^{1/2}}{(1-2|\varepsilon_0|(-1)^\mu)(1+2|\varepsilon_0|(-1)^v)}\right) + \left(\frac{(v+1)^{1/2}(\mu+1)^{1/2}}{(-1-2|\varepsilon_0|(-1)^\mu)(-1+2|\varepsilon_0|(-1)^v)}\right)\right] \cdot \delta_{v\mu}$$
$$+ \tfrac{1}{2}\lambda_0^2 \left[\left(\frac{(v)^{1/2}(\mu+1)^{1/2}\delta_{\mu+1,v-1}}{(1-2|\varepsilon_0|(-1)^\mu)(-1+2|\varepsilon_0|(-1)^v)}\right) + \left(\frac{(\mu)^{1/2}(v+1)^{1/2}\delta_{\mu-1,v+1}}{(-1-2|\varepsilon_0|(-1)^\mu)(1+2|\varepsilon_0|(-1)^v)}\right)\right] +$$
$$\tfrac{1}{\sqrt{2}}\lambda_0 \left\{\frac{(v^{1/2}\,\delta_{\mu,v-1})}{(1-2(-1)^v|\varepsilon_0|)} + \frac{((v+1)^{1/2}\,\delta_{\mu,v+1})}{(-1-2(-1)^v|\varepsilon_0|)} + \frac{(\mu^{1/2}\,\delta_{v,\mu-1})}{(1+2(-1)^\mu|\varepsilon_0|)} + \frac{((\mu+1)^{1/2}\,\delta_{v,\mu+1})}{(-1+2(-1)^\mu|\varepsilon_0|)}\right\} \quad (13)$$

Note that the transition dipole moment in eq. (13) is strongly dependent on the size of $\lambda_0$ and the selection rules as required by the Delta functions.

The second-order energy can be shown to be [20, 33]:

$$E_\pm^{(2)}(\mu) = \mu + \tfrac{1}{2} \pm (-1)^\mu|\varepsilon_0| \mp \lambda_0^2[(2\mu+1)(-1)^\mu|\varepsilon_0|/(1-4|\varepsilon_0|^2)] - \tfrac{1}{2}\lambda_0^2(1-4|\varepsilon_0|^2) \quad (14)$$

Note that $|\mu_+^{(1)}>$ and $|\mu_-^{(1)}>$ are associated with energies $E_+^{(1)}$ and $E_-^{(1)}$ respectively.

From (14) we can obtain the transition energy:

$$w_{IT}/w_0 = E_+^{(2)}(\mu) - E_-^{(2)}(v) = (\mu - v) + [1+(-1)^{v-\mu}]\cdot(-1)^\mu|\varepsilon_0| - \lambda_0^2\cdot|\varepsilon_0|\{[(2\mu+1)(-1)^\mu + (2v+1)(-1)^v]/(1-4|\varepsilon_0|^2)\} \quad (15)$$

From (15), when $\mu = v = 0$, at T= 0K the CT band's transition energy is [20]:

$$\mathcal{W}_{CT}/\mathcal{W}_0 = 2|\varepsilon_0| + 2\lambda_0^2/(4|\varepsilon_0|^2 - 1) \approx 2|\varepsilon_0| + \lambda_0^2/2|\varepsilon_0| \quad (16)$$

Equation 16 refers to the transition energy between the adiabatic states, i.e., from the lowest vibrational state in the lower potential surface $\mu = 0$ to the first allowed vibrational state $v = 0$ of the upper potential surface.

Note that the $v = 0$ to $\mu = 1$ transition is forbidden from (13). However the $v = 0$ to $\mu = 2$ is allowed with transition energy $2 + 2|\varepsilon_0| + 6\lambda^2\varepsilon_0/(4\varepsilon_0^2 - 1)$; and from (15) its intensity *is* $\lambda^2/2(1-2\varepsilon_0)^2$.

For the IR charge-transfer-induced transition, its energy is also obtained from (15):

$$E_+^{(2)}(1) - E_-^{(2)}(0) = 1 - 2|\varepsilon_0|\lambda_0^2/(4|\varepsilon_0|^2 - 1) \quad (17)$$

From (13), the inter-valence charge transfer transition intensity is:

$$I_{CT} = \left|\left\langle 0_-^{(1)}\middle|\delta n\middle|0_+^{(1)}\right\rangle\right|^2 = \left\{1 - \tfrac{1}{2}\lambda_0^2/(4|\varepsilon_0|^2 - 1)\right\}^2 \quad (18)$$

To obtain the charge-transfer-induced IR transition intensities, we use (13):

$$\left\langle \mu_+^{(1)} \middle| \delta n \middle| (\mu+1)_-^{(1)} \right\rangle = -\tfrac{1}{\sqrt{2}} \lambda_0 (-1)^\mu 4|\varepsilon_0|(\mu+1)^{1/2} / (4|\varepsilon_0|^2 - 1) \tag{19}$$

Setting $\mu = 0$ in (19), we get

$$I_{IR} = \left| \left\langle 0_+^{(1)} \middle| \delta n \middle| 1_-^{(1)} \right\rangle \right|^2 = \{8\lambda_0^2 \cdot |\varepsilon_0|^2 / (4|\varepsilon_0|^2 - 1)^2\} \tag{20}$$

The intensity ratio **R** = $I_{IR} / I_{CT}$ is

$$R = 8\lambda_0^2 \cdot |\varepsilon_0|^2 / (4|\varepsilon_0|^2 - 1 - \tfrac{1}{2}\lambda_0^2)^2 \approx \tfrac{1}{2}\lambda_0^2 / |\varepsilon_0|^2 \tag{21}$$

It should be emphasized that these transition energies and their corresponding intensities are valid for very low temperature for which kT is much less than the vibration frequency which is in the range of hundreds of cm$^{-1}$.

In expression (5) for the frequency-dependent conductivity, a Lorentzian line shape function is used [19].

The term $2w_{\beta 0}/[w_{\beta 0}^2 - (w+i\delta)^2]$ can be readily derived from (replace $\delta$ by $\eta$):

$$\left\{ \frac{1}{(w - w_{\beta 0} + i\eta)} - \frac{1}{(w + w_{\beta 0} + i\eta)} \right\}$$

$$= 4ww_{\beta o}\eta / [(w^2 - w_{\beta o}^2)^2 + \eta^4 + 2\eta^2(w^2 + w_{\beta o}^2)] \tag{22}$$

Note that the Lorentzian line shape function has a width of $2\eta$ at half height of the absorption maximum. In view of mixed valence absorption bands are mostly broad and featureless, for each allowed transition the value of $\eta$ is selected to be quite large for fitting the observed band contour. Of course there are other mechanisms to cause a broad contour such as coupling to some symmetric vibration modes [24], solvent effect [36], and the size of $\lambda_0$ as in Class I, etc.

The imaginary part of $X(w)$ gives the CT band which can be shown to be (with $w_0$ replaced by $w_{\beta o}$):

$$CT(w) = K_o 4M^2 w w_{\beta o} \eta / [(w^2 - w_{\beta o}^2)^2 + \eta^4 + 2\eta^2(w^2 + w_{\beta o}^2)] \tag{23}$$

Where $K_o = \tfrac{1}{4} e^2 a^2 N/\Omega$, $e$ and $a$ are defined in eq. (3), and $N$ is the number of dimers in volume $\Omega$.

Where $M^2 = |<\beta | \delta_n | 0 >|^2$ is the transition dipole moment,
$<\beta|$ denotes the eigenstate, and $|0>$ denotes the ground state.

Define $\eta = 2\gamma_m$

$$CT(w) = 8K_o M^2 w w_{\beta o} \gamma_m / [(w^2 - w_{\beta o}^2)^2 + 16\gamma_m^4 + 8\gamma_m^2(w^2 + w_{\beta o}^2)] \tag{24}$$

At T= 4.2 K, the only allowed transitions are from $v = 0$ to $\mu = 0$ & 2 as described earlier.

To calculate the CT band, we use the following for $w_{\beta o}$ (from eq. (16)):

$$w_{\beta o} = 2\varepsilon_0 + 2\lambda_0^2 \varepsilon_0/(4\varepsilon_0^2 - 1) \tag{25}$$

With intensity or transition dipole moment (from eq. 18):

$$M^2 = \{1 - \lambda_0^2/2(4\varepsilon_0^2 - 1)\}^2 \tag{26}$$

The optical conductivity due to CT band is given by

$$\sigma(\omega)_{CT} = w 8 K_o M^2 w w_{\beta o} \gamma_m / [(w^2 - w_{\beta o}^2)^2 + 16\gamma_m^4 + 8\gamma_m^2 (w^2 + w_{\beta o}^2)] \tag{27}$$

The IR band is given by the expression:

$$IR(w) = K_o M^2 [(8\lambda_0^2 |\varepsilon_0|^2/(4|\varepsilon_0|^2 - 1)^2](8 w w_\alpha \gamma_v)/[(w^2 - w_\alpha^2)^2 + 16\gamma_v^4 + 8\gamma_v^2 (w^2 + w_\alpha^2)] \tag{28}$$

$$\approx K_o M^2 [\lambda_0^2/2|\varepsilon_0|^2](8 w w_\alpha \gamma_v)/[(w^2 - w_\alpha^2)^2 + 16\gamma_v^4 + 8\gamma_v^2 (w^2 + w_\alpha^2)] \tag{29}$$

Where the first term in square bracket is the IR transition dipole moment from eq. 20. $w_\alpha$ is the anti-symmetric vibrational mode of the dimer, and $\gamma_v = \gamma_m$.

The optical conductivity due to IR band is given by

$$\sigma(\omega)_{IR} = w K_o M^2 [\lambda_0^2/2|\varepsilon_0|^2] (8 w w_\alpha \gamma_v)/[(w^2 - w_\alpha^2)^2 + 16\gamma_v^4 + 8\gamma_v^2 (w^2 + w_\alpha^2)] \tag{30}$$

Note that both $\sigma(\omega)_{CT}$ & $\sigma(\omega)_{IR}$ have the same polarization as the corresponding CT and IR bands respectively [17].

In addition the ratio of $\sigma(\omega)_{IR}/\sigma(\omega)_{CT}$ is proportional to $[\lambda_0^2/2|\varepsilon_0|^2](w_\alpha \gamma_v / w_{\beta o} \gamma_m)$. If the ratio $(\gamma_v/\gamma_m)$ is kept constant, and $[\lambda_0^2/2|\varepsilon_0|^2]$ is unchanged, then we have $w_\alpha/w_{\beta o} \ll 1$ when the CT band is located far from the IR band. Hence if $w_{\beta o}$ moves closer to $w_\alpha$, then $\sigma(\omega)_{IR}$ tends to increase its borrowed intensity.

### Delocalized Approach

The delocalized CT and IR bands' absorption profiles are calculated using expressions (24) and (29) respectively, with the aid of their respective transition energies and intensities from (16), (18), and (17), (20).

Figures 1 through 4 present the absorption profiles for a strongly delocalized case where $\lambda_0$ is nearly zero, and some not so delocalized cases with increasing $\lambda_0$ values. Note that the $K_o$ term is purposely omitted in these figures.

Once these absorption profiles are calculated, it is straightforward to calculate their respective optical conductivities $\sigma(\omega)_{CT}$ and $\sigma(\omega)_{IR}$. Note that for the strongly delocalized case, the intense CT band contributes to an intense optical conductivity band, with negligible contribution from the IR band. From Figures 1 through 4, it is evident that the $\sigma(\omega)$ profile is closely resemble that of the absorption profile, except that the peak of $\sigma(\omega)$ *always* shifts to higher frequency with respect to the corresponding absorption peak.

### LOCALIZED – CLASS I ($|\varepsilon_0| \ll \lambda_0^2$)

The vibronic Hamiltonian can be written as follows:

By applying the transformation to eq. 19, we have

$$sH_{vib}\,s^{-1} = H'_{vib}\,/\,\hbar w_\alpha = \begin{bmatrix} \frac{1}{2}(\dot{q}_0^2 + q_0^2) + \lambda_0 q_0 & \varepsilon_0 \\ \varepsilon_0 & \frac{1}{2}(\dot{q}_0^2 + q_0^2) - \lambda_0 q_0 \end{bmatrix} \quad (31)$$

where $s = \frac{1}{\sqrt{2}}\begin{pmatrix} 1 & 1 \\ 1 & -1 \end{pmatrix}$

The corresponding eigenstates are

$$s \begin{pmatrix} 1/\sqrt{2}\,(|1\rangle + |2\rangle) \\ 1/\sqrt{2}\,(|1\rangle - |2\rangle) \end{pmatrix} = \begin{pmatrix} |1\rangle \\ |2\rangle \end{pmatrix} \quad (32)$$

The above $H'_{vib}\,/\,\hbar w_\alpha$ can be written as

$$H'_{vib}/\hbar w_\alpha = \begin{bmatrix} \frac{1}{2}(\dot{q}_0^2) + \frac{1}{2}(q_0 + \lambda_0)^2 - \frac{1}{2}\lambda_0^2 & \varepsilon_0 \\ \varepsilon_0 & \frac{1}{2}(\dot{q}_0^2) + \frac{1}{2}(q_0 - \lambda_0)^2 - \frac{1}{2}\lambda_0^2 \end{bmatrix} \quad (33)$$

The secular determinant is [20]:

$$|H_{mn} - \delta_{mn}E| = 0 \quad (34)$$

Where

$$H_{mn} = (m + \frac{1}{2} - \frac{1}{2}\lambda_0^2)\,\delta_{mn} + \varepsilon_0 G_{mn} \quad (35)$$

where $G_{mn}\,(2\lambda_0)$ is the overlap integral between the vibrational wave functions given by

$$G_{mn}\,(2\lambda_0) = <x_m\,(q_0 - \lambda_0)|x_n\,(q_0 + \lambda_0)>$$
$$= (-1)^m e^{-\lambda_0^2}\,(2^{1/2}\lambda_0)^{n-m}\,(\frac{m!}{n!})^{1/2}\,L_m^{n-m}(2\lambda_0^2) \quad (36)$$

where $L_m^{n-m}(2\lambda_0^2)$ is the Laguerre polynomial.

The zeroth-order wave functions are [20]:

$$\Psi_\pm^{(0)}(\mu) = \tfrac{1}{\sqrt{2}}\{|1> x_\mu (q_0 + \lambda_0) + (-1)^\mu|2> x_\mu (q_0 - \lambda_0)\} \qquad (37)$$

And the first order energies are:

$$E_\pm^{(1)}(\mu) = \mu + 1/2 - 1/2\,\lambda_0^2 \pm (-1)^\mu \varepsilon_0 G_{\mu\mu}(2\lambda_0) \qquad (38)$$

The first-order wave function can be shown to be

$$\Psi_\pm^{(1)}(\mu) = \tfrac{1}{\sqrt{2}}\{|1> x_\mu (q_0 + \lambda_0) + (-1)^\mu|2> x_\mu (q_0 - \lambda_0)\} \mp$$

$$\tfrac{1}{\sqrt{2}}(-1)^\mu |\varepsilon_0|\{ |1> \sum_{\mu\mu'} \frac{G_{\mu\mu'}}{(\mu-\mu')} x_{\mu'}(q_0 + \lambda_0) + (-1)^\mu \sum_{\mu\mu''} \frac{G_{\mu\mu''}}{(\mu-\mu'')} |2> x_{\mu''}(q_0 - \lambda_0)\} \qquad (39)$$

Using the first-order wave function, the transition dipole moment is

$$|\langle\mu|\delta_n|0\rangle| = \left|\left\langle\Psi_\mp^{(1)}(\mu)\middle|\delta_n\middle|\Psi_\pm^{(1)}(0)\right\rangle\right|$$

$$= 2|\varepsilon_0| \cdot (2\lambda_0^2)^{\mu/2} \cdot e^{-\lambda_0^2}/(\mu^2 \cdot \mu!)^{1/2} \qquad (40)$$

The transition energies are given by

$$w_{\beta o} = w_{\mu o} = \left[E_+^{(1)}(\mu) - E_-^{(1)}(0)\right]$$

$$= \{\mu + |\varepsilon_0| \cdot [(-1)^\mu G_{\mu\mu}(2\lambda_0) - G_{00}(2\lambda_0)]$$

$$= \{\mu + |\varepsilon_0| \cdot [(-1)^\mu e^{-\lambda_0^2} \cdot L_\mu^0(2\lambda_0^2) - (-e^{-\lambda_0^2} \cdot L_0^0(2\lambda_0^2))]$$

$$= \{\mu + |\varepsilon_0| \cdot e^{-\lambda_0^2}[(-1)^\mu L_\mu^0(2\lambda_0^2) + 1]\} \qquad (41)$$

Note that when $|\varepsilon_0| = 0$, the transition energy is $\mu$, and $\boldsymbol{L_\mu^0}(2\lambda_0^2)$ or $\boldsymbol{L_\mu}(2\lambda_0^2)$ is given by [21].

Moreover, from (41) and [21], it is seen that there are many allowed transition energies from the ground state $E_-^{(1)}(0)$ (i.e., as many possible values of $\boldsymbol{L_\mu^0}(2\lambda_0^2)$ where $\mu \geq 1$. For illustration purpose, $\mu \leq 30$ is assumed in this paper and this will restrict $\lambda_0$ to be less than 4 for getting a reasonable absorption profile.

### Localized Approach
The localized CT band's absorption profile is calculated using expressions (24), with the aid of their respective transition energies and intensities from (41) and (40). In calculating the total absorption profile, a component profile for each allowed transition with energy $w_{\mu o}$ and its corresponding intensity is evaluated using (24). The total profile is then the sum of these component profiles (30 are used here). The total $\sigma(\omega)_{CT}$ profile is likewise calculated as the sum of the component $\sigma(\omega)_{CT}$ profiles using (27).

Figures 5 through 7 present the absorption profiles for a strongly localized case where $|\varepsilon_0|$ is very small, and some not so localized cases with increasing $|\varepsilon_0|$ values. Note that the $K_o$ term is purposely omitted in these figures. It is seen that the CT band absorption and optical conductivity band profiles are much larger compared to that for the Delocalized cases in Figures 1 through 4.

Note that for the strongly localized case, the weak CT band contributes to similarly weak optical conductivity band. There is no IR band and hence no contribution to the optical conductivity. From Figures 5 through 7, it is evident that the $\sigma(\omega)_{CT}$ profile is closely resemble that of the absorption profile, except that the peak of $\sigma(\omega)_{CT}$ *always* shifts to higher frequency with respect to the corresponding absorption peak.

In the localized case, a potential barrier (assume $hw_\alpha = 1$) $E_a = \frac{1}{2}\lambda_0^2 - |\varepsilon_0| + \frac{|\varepsilon_0|^2}{2\lambda_0^2}$ appears with activation energy $E_b = \frac{1}{2}\lambda_0^2$ [22]. The electron tunneling transfer rate through the barrier from one potential well to the other can be calculated using Weiner's method, and the result is (when T → 0 K):

$$\bar{P} = 4\pi^2 w_\alpha |\varepsilon_0|^2 \, e^{-2\lambda_0^2} \, [L_0^0(2\lambda_0^2)]^2 = 4\pi^2 w_\alpha |\varepsilon_0|^2 \, e^{-2\lambda_0^2} \tag{42}$$

This is exactly the same result obtained by using the Frank-Condon approximation.

However, since there is a dynamic thermal equilibrium, the tunneling transfer rate applies to *both* directions and hence the *net* transfer rate is zero. This result is contrary to that suggested by Hush [23] that there is a non-zero optical conductivity $\sigma_0$ at *zero frequency* caused by the electron tunneling transfer rate (i.e., also called "thermal rate" by Hush).

Note that the electron tunneling rate does contribute to the DC conductivity in the presence of an external *static* electric field $E_s$ parallel to the chain of donor-acceptor ions/moiety. The mobility $\mu$ is given by

$$\mu = a \cdot \text{net transfer rate}/E_s \ = [4a\pi^2 w_\alpha |\varepsilon_0|^2 e^{-2\lambda_0^2}]/E_s \tag{43}$$

Hence $\sigma(T\to 0) = ne\mu/E_s \ = [4nea\pi^2 w_\alpha |\boldsymbol{\varepsilon_0}|^2 \boldsymbol{e}^{-2\lambda_0^2}]/E_s \tag{44}$

Where n is the number of donor-acceptor ions/moiety per unit volume and e is the electron charge. This is exactly the same result that can be derived using Jortner's expression for electron transfer rate [13].

**INTERMEDIATE – CLASS II** ($|\varepsilon_0| \approx \lambda_0^2$)

Figures 8 through 11 present the absorption and optical conductivity profiles calculated via the delocalized and localized approaches for small values of both $|\varepsilon_0|$ and $\lambda_0$. Two sets of parameters are presented: $|\varepsilon_0| = \lambda_0 = 0.4$ and $1.0$.

Note that for the delocalized approach the following are observed:

1. The absorption band profile is contributed by both the CT band and the IR band.

2. Both the CT band and the IR band are in close proximity and have comparable intensities. For $|\varepsilon_0| = \lambda_0 = 0.4$ case, it is seen that the IR band is stronger than the CT band.

3. Proximity of CT and IR bands both increase the optical conductivity.

4. The calculated peak of $\sigma(\omega)$ profile always shifts to higher frequency with respect to the corresponding absorption peak.

The localized approach gives very similar absorption profile and charge transfer absorption profile in comparison with the delocalized approach, except with less intensities and slightly different peak positions. This is expected as the IR band contribution is absent as discussed above.

In summary the calculated absorption profiles are using the Lorentzian line shape function whose bandwidth $\eta$ is assigned a fixed value, similar to the Gaussian bandwidth in the PKS model. The value of $\eta$ for the IR band is assumed to be the same as that of the CT band, as done in the PKS model.

Moreover, the band shape of the charge transfer transition is determined by the size of $\lambda_0$ and the mode of vibrations. Indeed, it is true that IR band should have a very sharp/narrow linewidth with its intensity borrowed from the CT band.

In a paper [24] it is shown that the symmetric mode vibration of the bridging site between the charge-related sites (e.g., Ru ions in Creutz & Taube complex) significantly contributes to the bandwidth of the CT band.

It is found that there for Class I system (strongly localized case) there is no optical conductivity due to electron tunneling transfer rate as frequency tends to zero.

CONCLUSION

The calculated absorption profiles for all the mixed-valence Classes are in fairly good agreement with that of the PKS model when the perturbation term is small with respect to the dominant term (e.g., $\lambda_0^2 \ll |\varepsilon_0|$ for the Delocalized case). This is expected as the vibronic Hamiltonian is solved by using perturbation approach.

Without the use of full diagonalization as in the PKS approach, the above approach provides another means of calculating the mixed-valence systems' absorption profiles and optical conductivity profiles. It is shown that the peaks of the latter are slightly shifted toward the higher frequency than the former in all mixed-valence Classes, and no contribution to optical conductivity from electron tunneling transfer in Class I (strongly localized) case.

Optical conductivity has been discussed extensively both theoretically [25-27] and experimentally [28, 29]. It has been used successfully in evaluating the physical properties and the related mechanisms in many systems [26, 28, 29], it is strongly suggested that it could be applicable to mixed-valence systems. The expression (27) can be applied to fit the measured optical conductivity at 4.2 K [30] based on its polarization, peak position, intensity and bandwidth to obtain the relevant $\varepsilon_0$ $and$ $\lambda_0$ .

## ACKNOWLEDGEMENT

The author is grateful to Professor Jeffrey Reimers of University of Technology Sydney for reviewing an earlier draft of this article and providing helpful suggestions and discussions.

## LIST OF REFERENCES

## Figure Captions

**Figure 1.** Delocalized case with $|\varepsilon| = 6.0$ and $\lambda = 0.01$. The dashed-curve is the absorption profile (right side scale), and the continuous cure is the optical conductivity profile (left side scale). Horizon axis is in units of $\hbar v_0$. These profiles are calculated with T= 0 K and with equal bandwidth $1.2\, v_0$ and $v_0$ is set to 1. Note that the absorption peaks at 12.0, which is in good agreement with that of PKS result of 12.0 [Figure 7 in ref. 22].

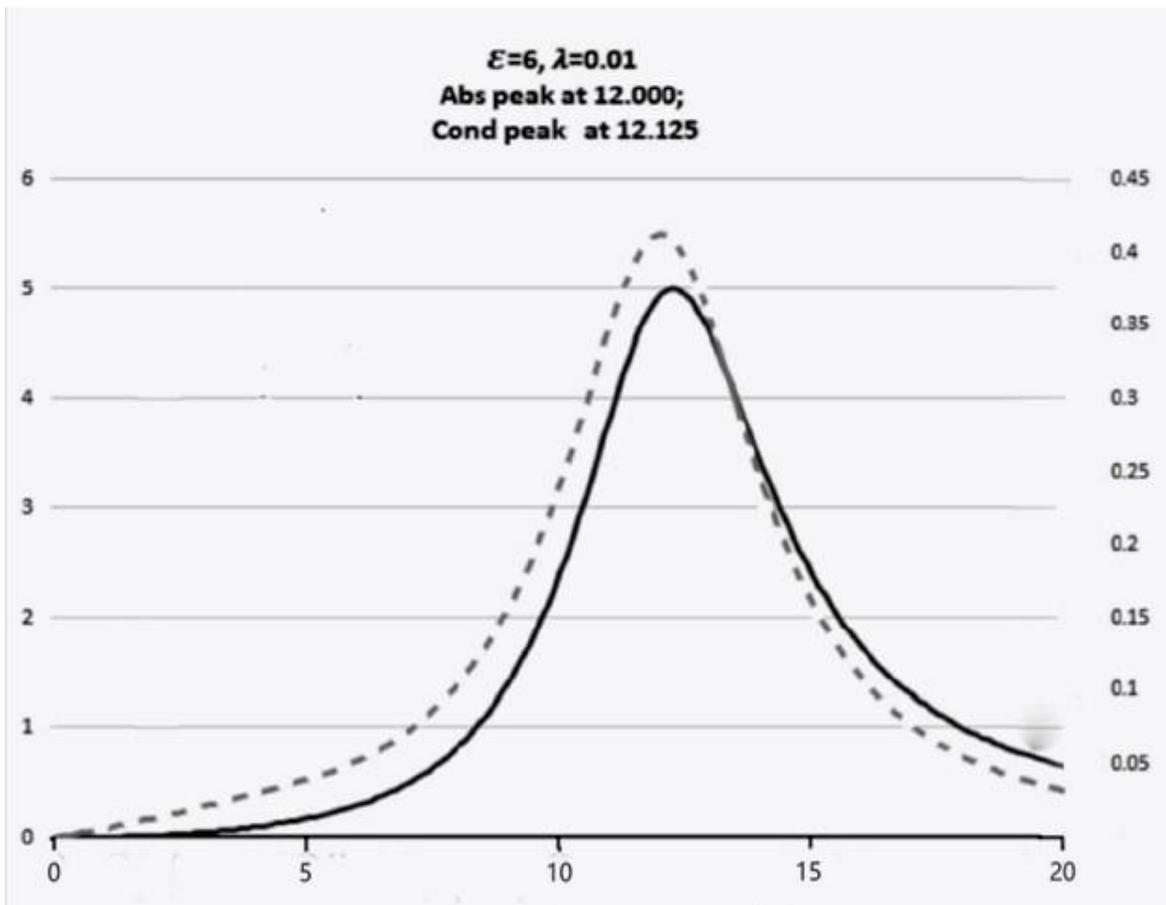

**Figure 2.** Delocalized case with $|\varepsilon| = 6.0$ and $\lambda = \sqrt{6} = 2.449$. Note that with large $\lambda$ value (in this particular case the lower potential surface is flat) and there is significant IR band contribution to the total absorption profile at low frequencies. Note that the absorption peaks at 12.5, which is in good agreement with that of PKS result [Figure 7 in ref. 22].

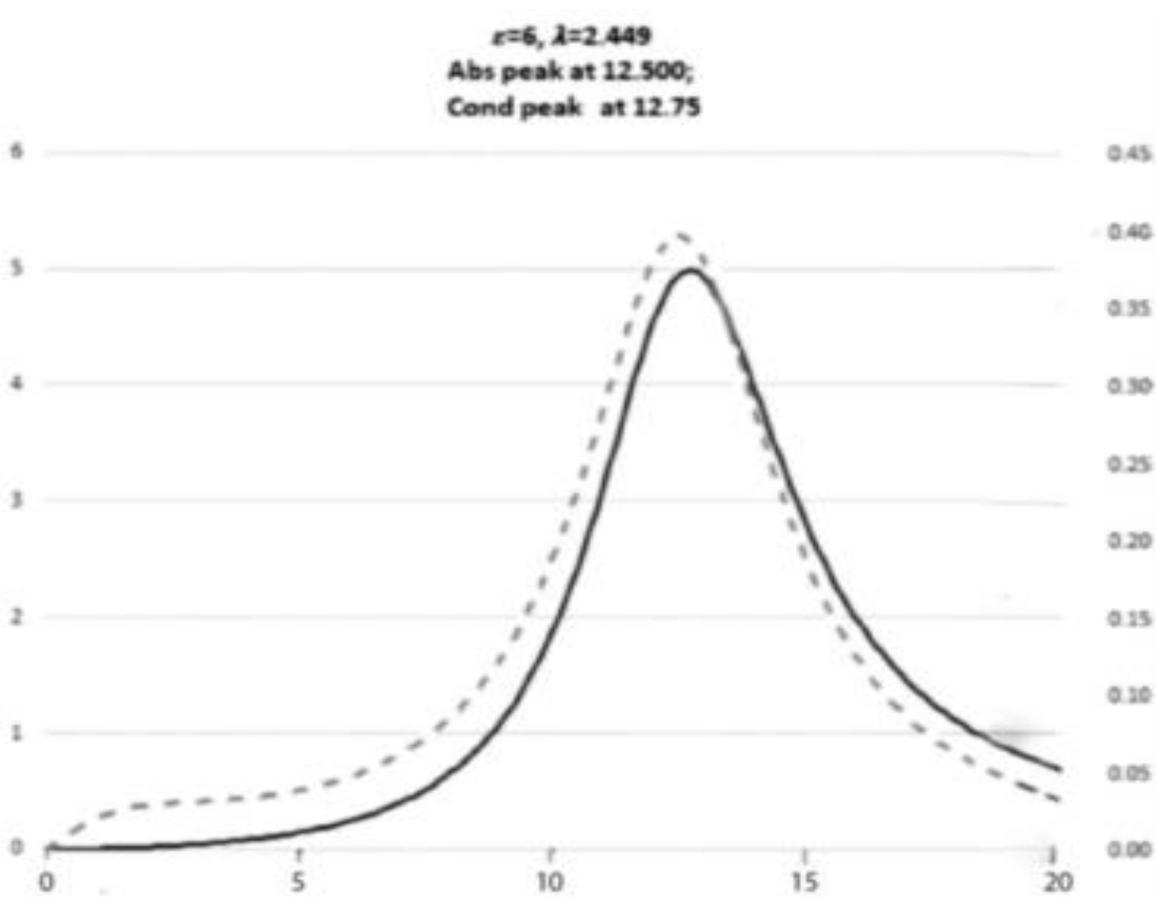

**Figure 3.** Delocalized case with $|\varepsilon| = 3.0$ and $\lambda = 0.1$. Note that the absorption peaks at 6.0, which is in good agreement with that of PKS result [Figure 7 in ref. 22].

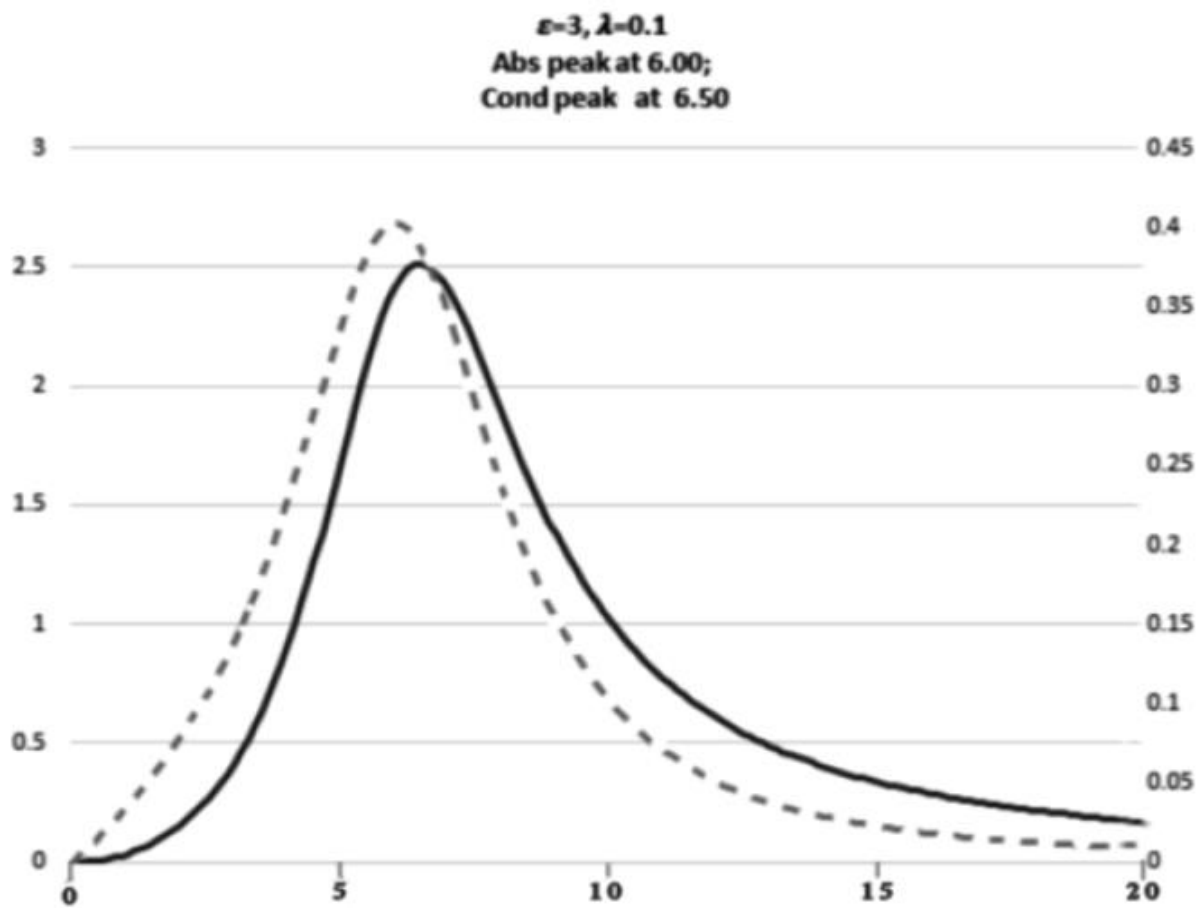

**Figure 4.** Delocalized case with $|\varepsilon| = 3.0$ and $\lambda = \sqrt{3} = 1.732$. Note that the IR band contribution to the absorption profile at low frequencies becomes more prominent when $|\varepsilon|$ is not much larger than $\lambda$. Note that the absorption peaks at 6.5, which is in fairly good agreement with that of PKS result of 6.8 [Figure 7 in ref. 22].

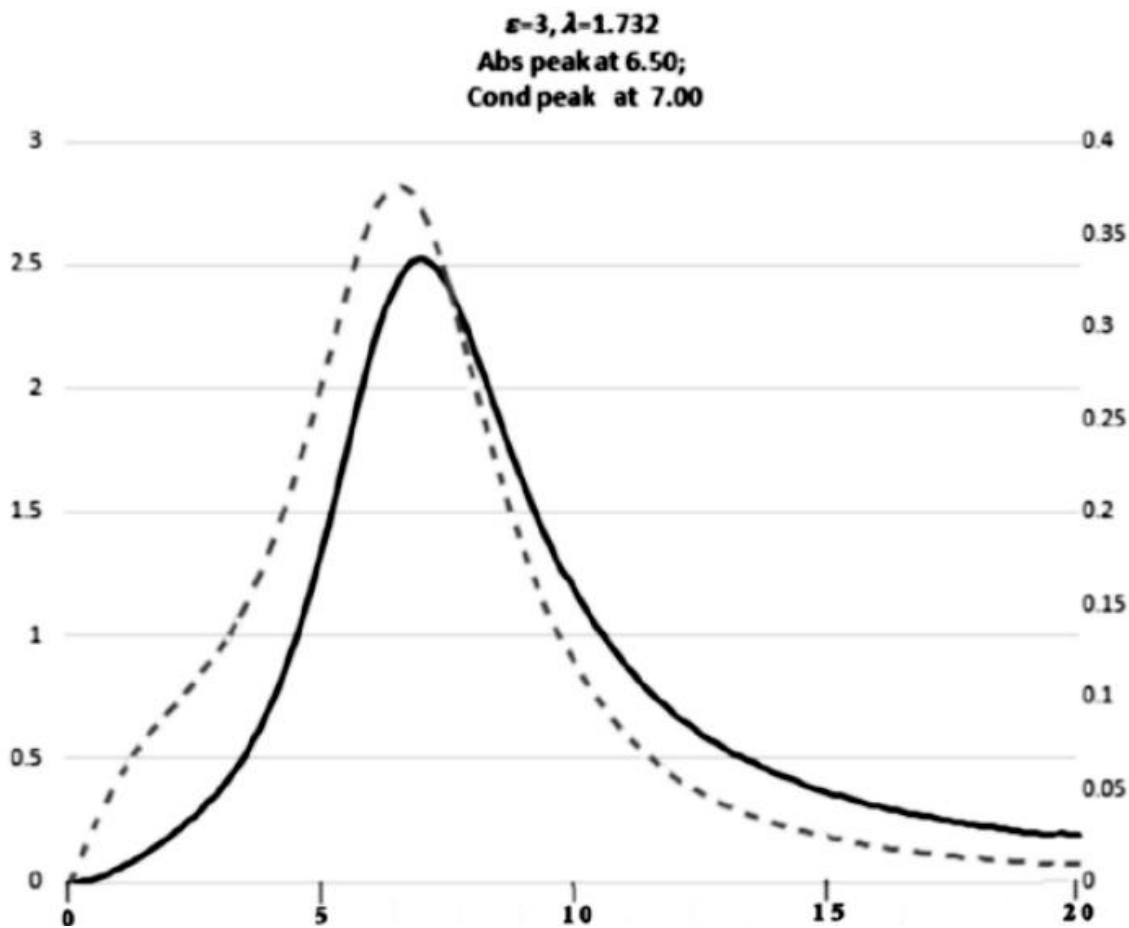

**Figure 5.** Localized case with and $\lambda = 3.0$ $|\varepsilon| = 0.01$. The dashed curve is the absorption profile (right side scale), and the continuous cure is the optical conductivity profile (left side scale). Each profile is calculated with T= 0 K and with equal bandwidth 1.2 $v_0$ and $v_0$ is set to 1. Note that the absorption peaks at 15.5, which is in fairly good agreement with that of PKS result of ~ 15.3 [Figure 7 in ref. 22].

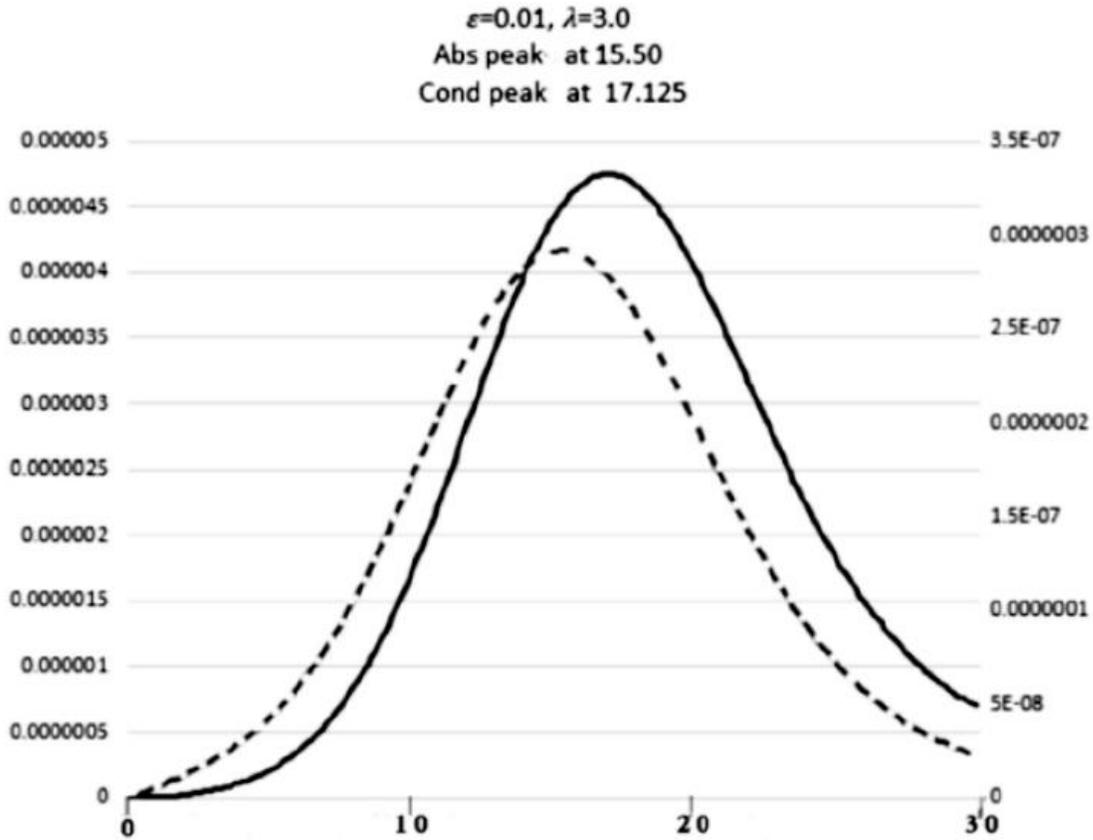

**Figure 6.** Localized case with and $\lambda = 3.0$ $|\varepsilon| = 1.0$. The dashed curve is the absorption profile (right side scale), and the continuous cure is the optical conductivity profile (left side scale).

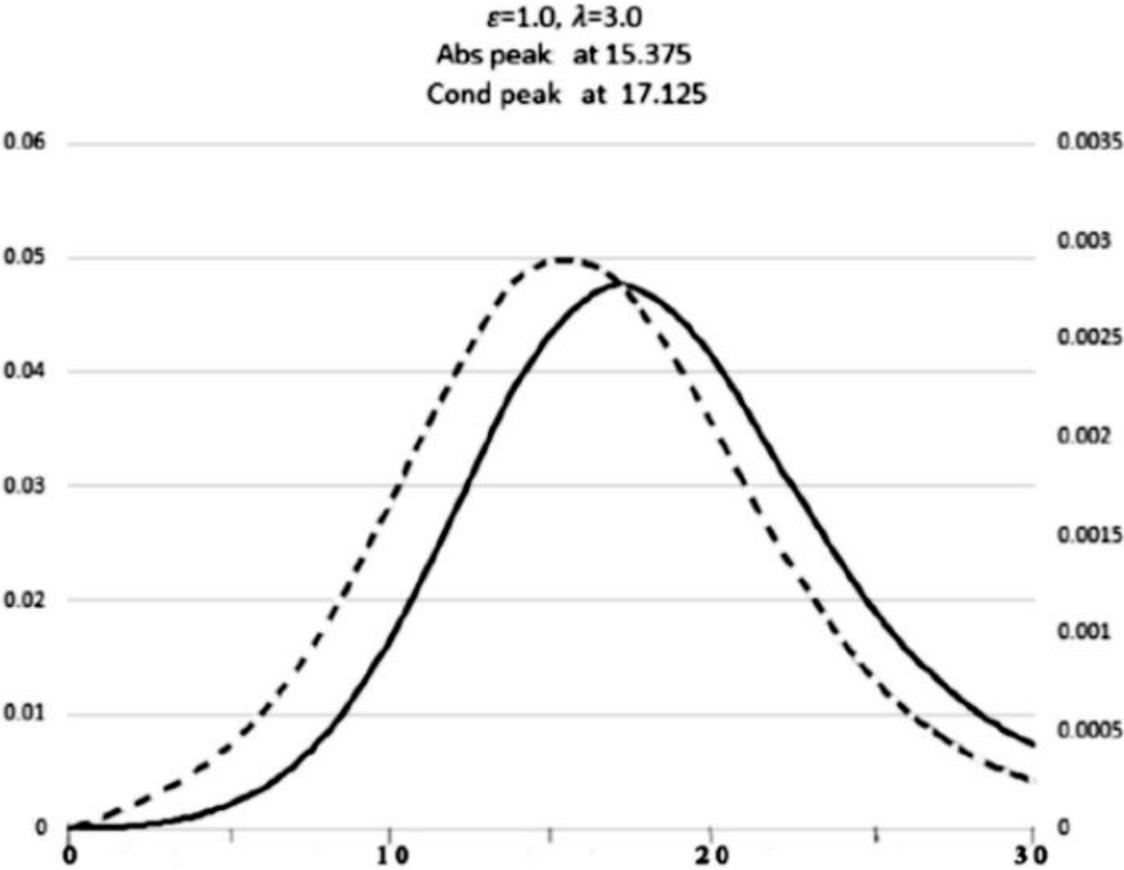

**Figure 7.** Localized case with and $\lambda = 3.0$ $|\varepsilon| = 2.0$ The dashed curve is the absorption profile (right side scale), and the continuous cure is the optical conductivity profile (left side scale). Note that the absorption peaks at 15.375, which is in good agreement with that of PKS result of ~15.5 [Figure 7 in ref. 22].

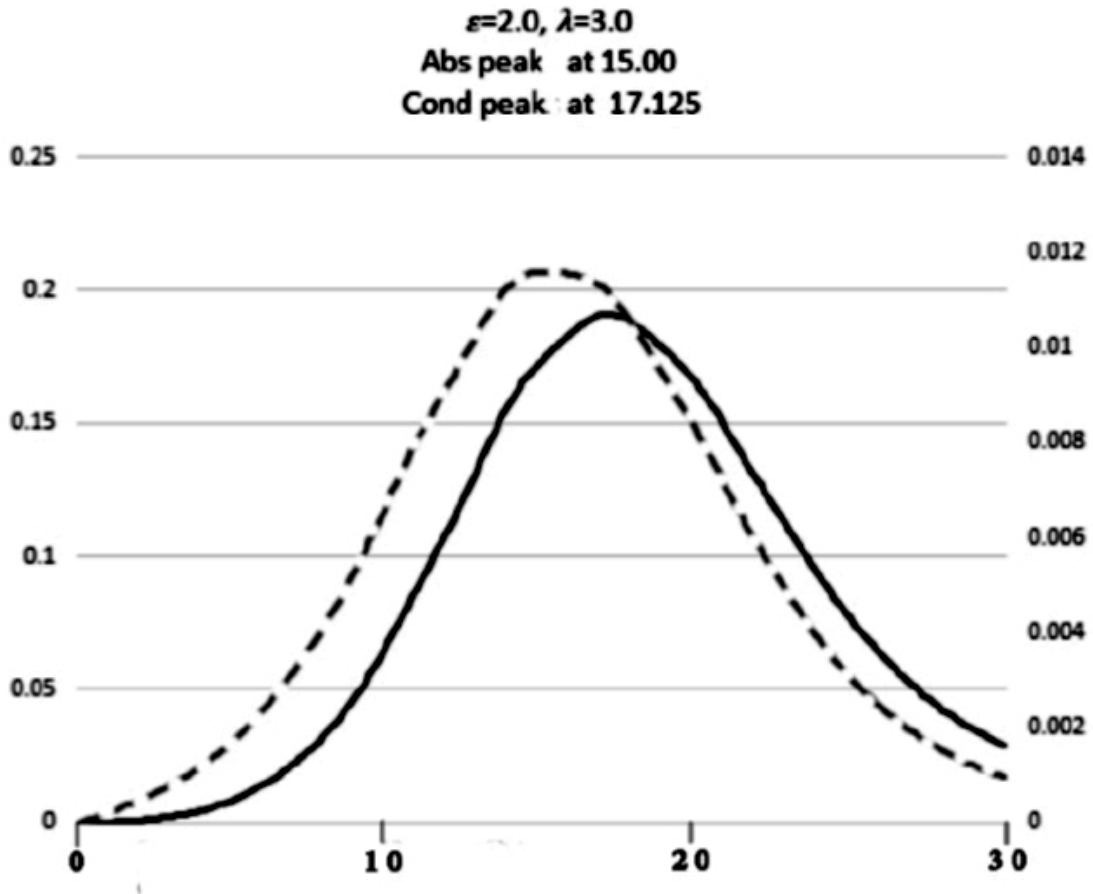

**Figure 8.** Delocalized approach to calculate profiles with $|\varepsilon| = \lambda = 0.4$. Note that the absorption profile (dashed curve, right-hand scale) has both contributions from the CT band (dotted curve, left-hand scale) and the IR band (dotted-dashed curve, left-hand scale), the latter is seen to be larger than the CT band contribution. The optical conductivity profile (continuous curve, left-hand scale) is shown shifted toward higher frequency with respect to the absorption profile. Each profile is calculated with T= 0 K and with equal bandwidth 1.2 $v_0$ and $v_0$ is set to 1. Note that the intensity of the IR band is almost twice that of the CT band.

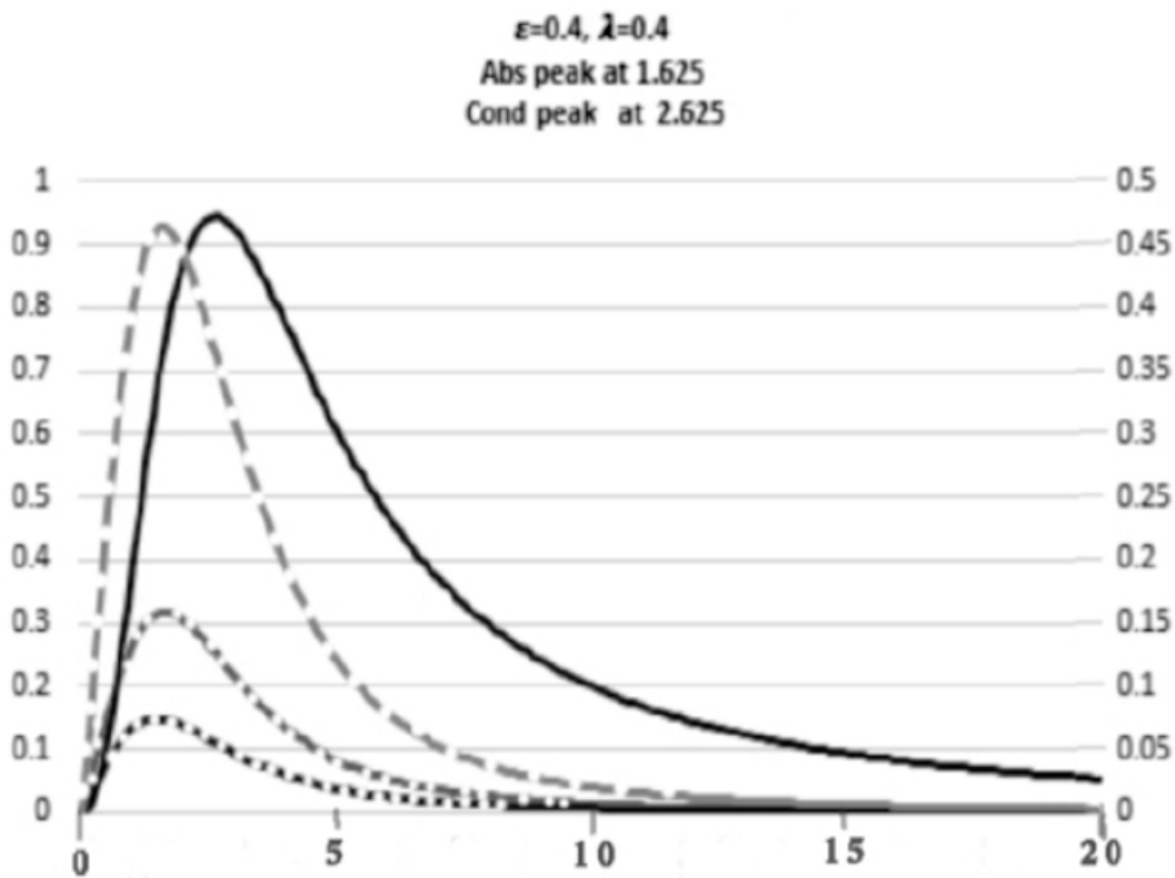

**Figure 9.** Localized approach to calculate profiles with $|\varepsilon| = \lambda = 0.4$. Note that only the absorption profile and its associated $\sigma(\omega)$ profile are presented here. There is no contribution to $\sigma(\omega)$ due to electron tunneling transfer rate as discussed above. Note that the intensities of the CT band and $\sigma(\omega)$ are order of magnitude less than the corresponding ones in the delocalized approach: 0.03 vs 0.46 for the CT band, and 0.067 vs 0.94 for $\sigma(\omega)$. This can be attributed to the omission of IR band contribution in the delocalized approach which is seen to be significant in the delocalized approach in Figure 8.

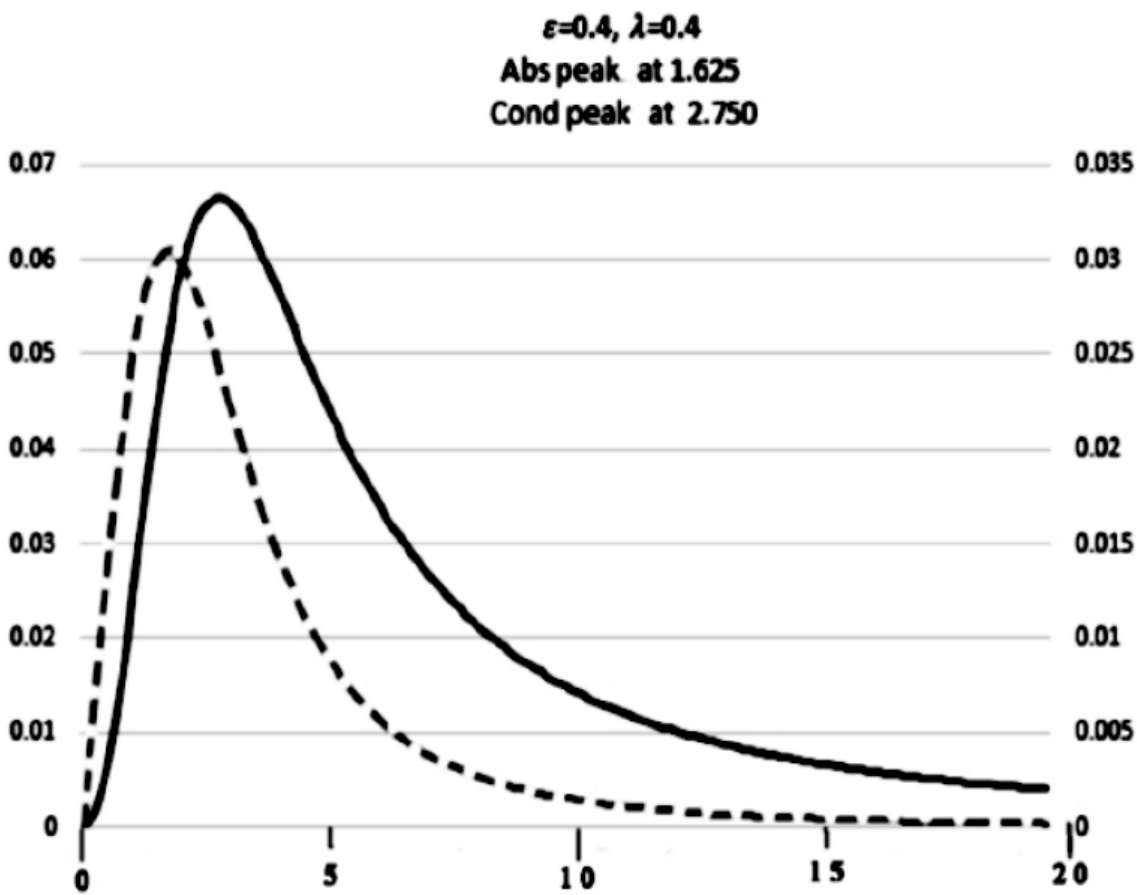

**Figure 10.** Delocalized approach to calculate profiles with $|\varepsilon| = \lambda = 1.0$. Note that the absorption profile (dashed curve, right-hand scale) has both contributions from the CT band (dotted curve, left-hand scale) and the IR band (dotted-dashed curve, left-hand scale), the former is seen to be larger than the IR band and is not close to the IR band. Note that the absorption peaks at 2.375, which is in good agreement with that of PKS result of ~2.45 [Figure 7 in ref. 22]. Moreover the delocalized approach is more compatible with the PKS result as they are able to include both the CT band and IR band contributions to the absorption profile.

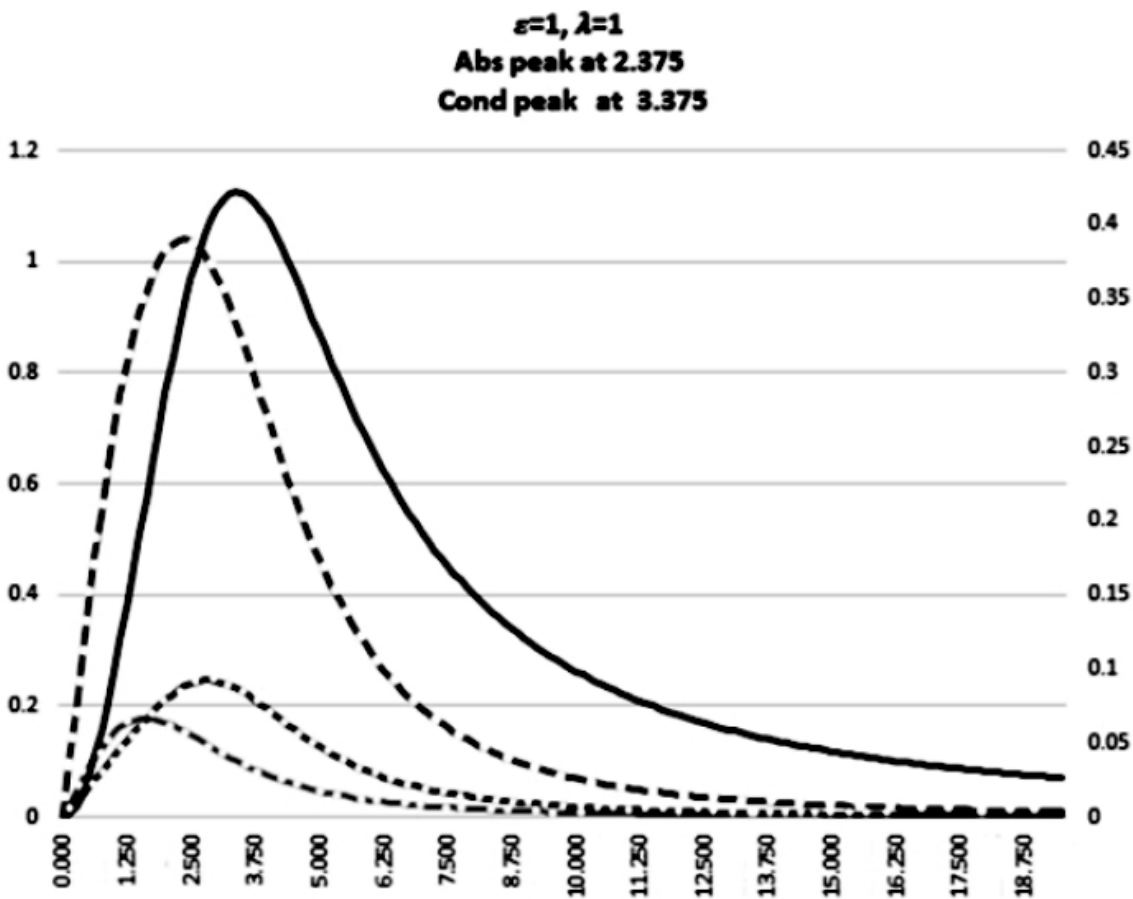

Fgure 11. Localized approach to calculate profiles with $|\varepsilon| = \lambda = 1.0$. Note that only the absorption profile (dashed curve, right-hand scale) and its associated $\sigma(\omega)$ profile (continuous curve, left-hand scale) are presented here. There is no contribution to $\sigma(\omega)$ due to electron tunneling transfer rate as discussed above. Note that the absorption peaks at 2.25, which is in good agreement with that of PKS result of ~2.45 [Figure 7 in ref. 22]. The difference in absorption peak position between the localized approach and the delocalized approach can be attributed to the presence of the IR band contribution for the latter which causes the peak moving to higher frequency than that of the localized approach. In addition, the intensities (peak heights) of the absorption and $\sigma(\omega)$ are ~70% of the corresponding intensities in the delocalized approach in Figure 10.

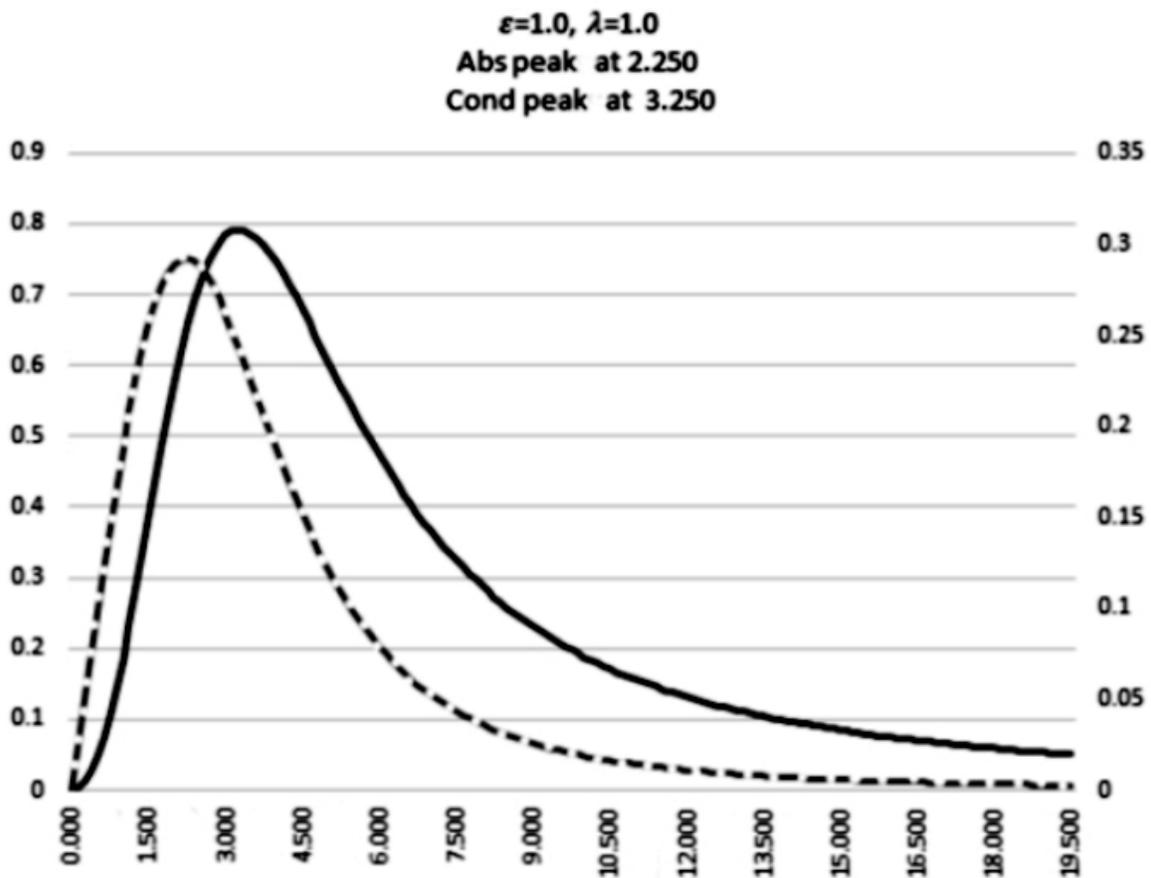